\documentclass[prl,twocolumn,superscriptaddress,showpacs]{revtex4}

\usepackage{graphicx}
\usepackage{txfonts}
\usepackage{bm}
\usepackage{color}
\usepackage{hyperref}

\begin{document}

\title{Structural Domain Walls in Polar Hexagonal Manganites}

\author{Yu Kumagai}
\email[]{yu.kumagai@mat.ethz.ch}
\author{Nicola A. Spaldin}
\affiliation{Department of Materials, ETH Zurich, Wolfgang-Pauli-Strasse, Z$\ddot{u}$rich 8093, Switzerland}
\date{\today}

\definecolor{blue}{rgb}{0.0,0.0,1}
\hypersetup{colorlinks,breaklinks,linkcolor=blue,urlcolor=blue,anchorcolor=blue,citecolor=blue}

\begin{abstract}
We present a microscopic first-principles study of the neutral structural domain walls (DWs) in the multiferroic 
hexagonal manganites, which have been shown to exhibit cross-couplings between ferroelectricity and structural antiphase.
We find that, in contradiction with previously proposed models, the interlocked antiphase-ferroelectric domain 
walls have approximately zero width, and their energy is lower than that of antiphase-only or ferroelectric-only domain walls.
Furthermore, we show that the ferroelectric-only and antiphase-only DWs are superpositions of interlocked antiphase ferroelectric (AP+FE) DWs and intermediate domains inevitably exist through the DWs.
Our results shed light on the question of why only AP+FE DWs are observed and how the topological defects emerge in polar hexagonal manganites.
\end{abstract}


\maketitle
In ferroic materials such as ferroelectrics, regions that differ in the orientation of the ferroic order
parameter (in ferroelectrics, the electrical polarization) are called domains. The planar defects between
adjacent domains are called domain walls (DWs). In general, DWs have a different symmetry and structure
than the bulk of the domains, and play an important role in the switching mechanism of the primary order
parameter. In multiferroic materials, which contain multiple simultaneous ferroic orders, the DWs are of
particular interest. For example, the ferroelectric DWs in multiferroics have been 
shown to exhibit novel properties such as local conduction~\cite{NatMat.8.229,PhysRevLett.107.127601,PhysRevLett.108.077203,NatMat.11.284} and ferromagnetism~\cite{Arxiv.1201.0694}. 
In addition, couplings among multiple (anti-)ferroic orders, that are not observed in the bulk
systesm, have been reported at the DWs ~\cite{Nature.419.818,NatMat.8.558,RevModPhys.84.119}. 

\begin{figure}
  \includegraphics[width=1\linewidth]{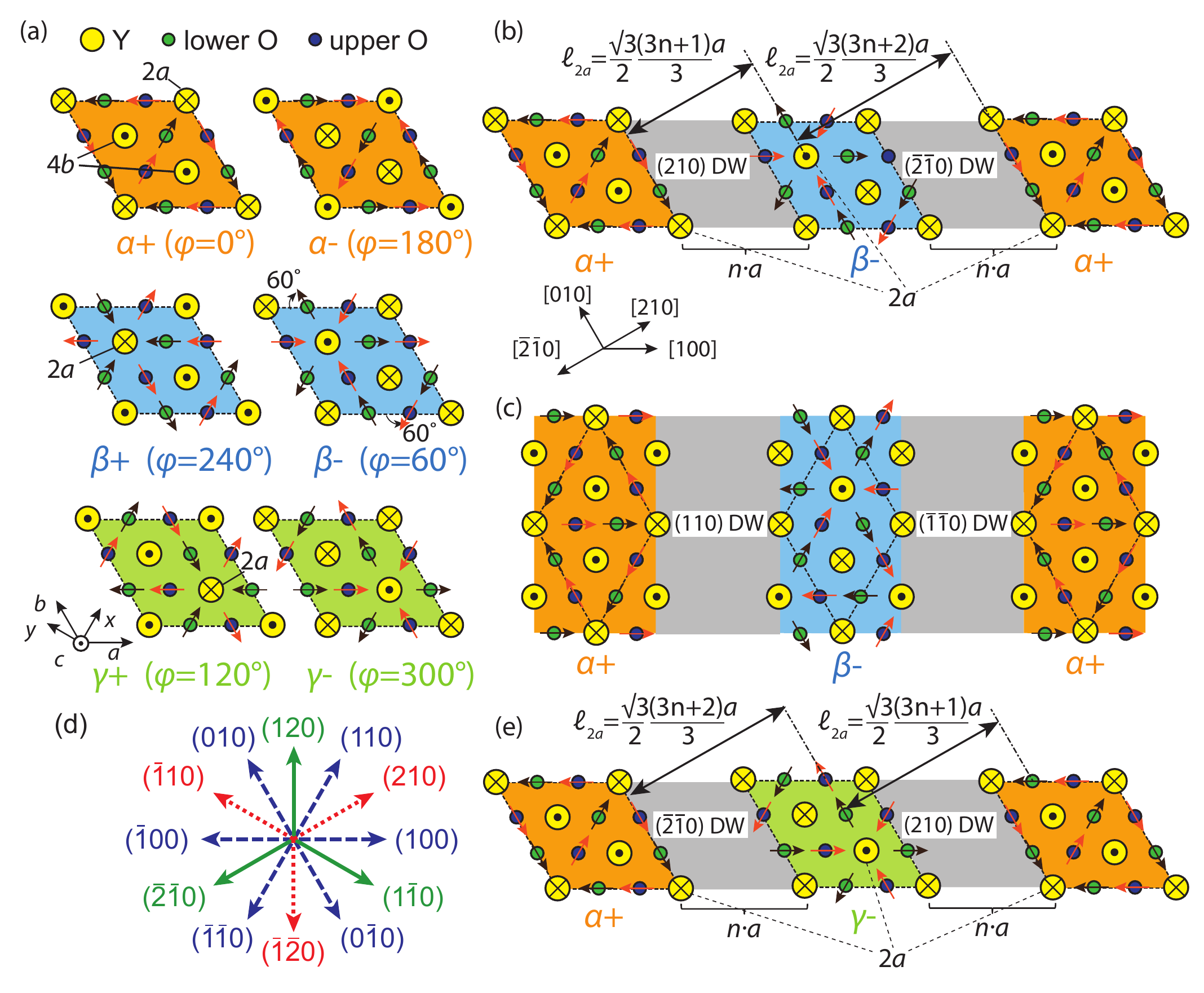}
  \caption{(a) Schematics of the 6 possible structural domains in YMnO$_3$. The view is along the $z$ axis, 
           perpendicular to the direction of ferroelectric polarization. 
           The big and small circles show respectively the positions of Y ions and adjacent oxygens at the apices of the trigonal bipyramids, 
           and their arrows indicate the direction of their displacements from the high symmetry $P6_3/mmc$ phase. 
           We label each domain by its phase ${\it \varphi}$, 
           defined to  be the (counter)clockwise angle of the displacement directions of upper (lower) oxygens relative to $\alpha^+$ domain~\cite{Arxiv.1204.4126}.
           Note that the $2a$ Wyckoff positions vary regarding the antiphase domains.
           (b--c) Our nomenclature for the neutral DWs sandwiched between $\alpha^+$ and  $\beta^-$ domains.
           Note that the $(210)$ and $(\bar{2}\bar{1}0)$ DWs are different whereas $(110)$ and $(\bar{1}\bar{1}0)$ DWs are same by symmetry (see text for details).
           (d) The classification of the equivalent DWs between $\alpha^+$ and  $\beta^-$ domains that are shown with normal vector directions.
           (e) DWs between $\alpha^+$ and $\gamma^-$ domains. 
           The $(210)$ and $(\bar{2}\bar{1}0)$ DWs are reversed from those between $\alpha^+$ and  $\beta^-$ domains.
          }
  \label{DW_definition}
\end{figure}

The hexagonal manganites h-$R$MnO$_3$ ($R$=Sc, Y, Dy--Lu), with their coexisting ferroelectricity 
($T_{\rm C}$$\approx$1200--1500~K) and antiferromagnetism ($T_{\rm N}$$\approx$70--130~K), are 
currently among the most intensely investigated 
multiferroics~\cite{Nature.419.818,Nature.430.541,Nature.451.805,NatMat.430.541}.
They are improper geometric ferroelectrics, in which the primary order parameter is a structural
tilt trimerization of the MnO$_5$ polyhedra, which is driven by minimization of the electrostatic 
potential~\cite{PhysRevB.72.100103,NatMat.430.541,PhysRevB.85.174422}.
The trimerizaton corresponds to the condensation of a zone-boundary $K_3$ mode of the high symmetry
$P6_3/mmc$ structure, and lowers the symmetry to the polar $P6_3cm$ space group~\cite{PhysRevB.72.100103}
but does not itself introduce a net ferroelectric polarization.
The trimerization can be described by two angles: the angle between the $z$ axis and the Mn-apical O bond
gives the magnitude of the tilting and can be regarded as the order parameter of the $K_3$ mode, and 
an azimuthal angle, $\varphi$, which describes the orientation of the tilting. Below the phase transition
$\varphi$ adopts one of six values separated by 60$^{\circ}$, which correspond to trimerization around
three possible origins, with in- or out- orientation of the tilting; as a result six structural domains
emerge, as shown in  Fig~\ref{DW_definition}(a)~\cite{Arxiv.1204.4126}. 
The polar $\Gamma^-_2$ mode emerges as a result of an anharmonic coupling with the $K_3$ mode, and the
orientation of the subsequent ferroelectric polarization is set by the in- or out- orientation of the
$K_3$ tilting.
Then, $\varphi=0^{\circ}$, 60$^{\circ}$, $120^{\circ}$, $180^{\circ}$, $240^{\circ}$, and $300^{\circ}$ 
are often labeled $\alpha^+$,
$\beta^-$, $\gamma^+$, $\alpha^-$, $\beta^+$, and $\gamma^-$ (Fig.~\ref{DW_definition}(a)), where 
$\alpha$, $\beta$ and $\gamma$ represent different origins for the trimerization, and $+$ or $-$ 
indicate the out- or in- orientation of the tilts and corrsepondingly the up- or down- orientation
of the ferroelectric polarization.

There are six domains in $R$MnO$_3$, and in principle, three types of structural DWs might be expected -- ferroelectric only (FE), antiphase only
(AP), and antiphase plus ferroelectric (AP+FE)  -- however, recent experiments showed that only AP+FE DWs exist, 
and that FE only DWs and AP only DWs do not occur~\cite{NatMat.9.253,ApplPhysLett.97.012904}.
A net ferromagnetic signal has been reported at the AP+FE walls~\cite{Arxiv.1201.0694}, 
as well as a wall-mediated coupling between the antiferromagnetic and ferroelectric domains~\cite{Nature.419.818}.
Finally, the intersections of the six possible AP+FE DWs form characteristic vortex defects
which have been shown to be topologically protected~\cite{NatMat.9.253,ApplPhysLett.97.012904,NatMat.11.284,PhysRevB.85.174422,Arxiv.1201.0694,Arxiv.1204.3785,PhysRevLett.108.167603}.

In this work we present a detailed systematic study of DW structures and energetics in YMnO$_3$ using first-principles
density functional theory. Our results are consistent with and complementary to a recent microscopy study using scanning 
transmission electron microscopy (STEM)~\cite{PhysRevB.85.020102}, which is not able to image the positions of light ions.
In particular, we are able to explain the origin of the peculiar AP+FE selection rule. 

We begin by introducing our nomencalutre for the neutral DWs in h-$R$MnO$_3$.
We first consider DWs between $\alpha^+$ and $\beta^-$ domains.
We define the $(210)$ DW to be the wall of which normal vector from an $\alpha^+$ domain to a $\beta^-$ domain points the [210] direction with respect to the primitive unit cell vectors 
(Fig.~\ref{DW_definition}(b), left wall).
The $(\bar{2}\bar{1}0)$ DW is then the wall with the same orientation, but the $\alpha^+$ and $\beta^-$ domains are reversed, as shown on the right of Fig.~\ref{DW_definition}(b). 
Although the detailed structures are not known yet, the structures of the $(210)$ and $(\bar{2}\bar{1}0)$ DWs are in fact different, as can be seen for example by comparing
the distances between the layers of $R$ ions staying at $2a$ Wyckoff position~\cite{PhysRevB.83.094111} in each domain, $\ell_{2a}$, as the wall is traversed. 
We see in Fig.~\ref{DW_definition}(b) that
$\ell_{2a}=\frac{\sqrt3}{2}\cdot\frac{(3n+1)}{3}a$ (where $n$ is an integer) across a $(210)$ DW and 
$\ell_{2a}=\frac{\sqrt3}{2}\cdot\frac{(3n+2)}{3}a$ across a $(\bar{2}\bar{1}0)$ DW. 
Since the geometries are different the structures and energetics must be in principle different. 
On the other hand, the $(\bar{1}\bar{2}0)$, $(\bar{1}10)$ and $(210)$ DWs 
(normal vector from $\alpha^+$ to $\beta^-$ pointing [$\bar{1}\bar{2}$0], [$\bar{1}$10] and [210] directions) 
are all equivalent to each other by symmetry, 
and the $(120)$, $(1\bar{1}0)$ and $(\bar{2}\bar{1}0)$ DWs (defined analogously) also form a symmetry-equivalent set.
We refer to these two types of DWs as $\{210\}$ and $\{120\}$ DWs respectively.
The two DWs perpendicular to the $[110]$ direction are $(110)$ and $(\bar{1}\bar{1}0)$ DWs (Fig.~\ref{DW_definition}(c)), 
and they are identical by symmetry.
$(100)$, $(010)$, $(\bar{1}00)$, and $(0\bar{1}0)$ DWs are also equivalent to $(110)$ and $(\bar{1}\bar{1}0)$ DWs, and we refer to 
this type of DW as $\{110\}$ DW.
The DWs between $\beta^+$ and $\gamma^-$, and between $\gamma^+$ and $\alpha^-$ are equivalent to that between $\alpha^+$ and $\beta^-$, for the same DW direction.
Furthermore, the DW between $\alpha^-$ and $\beta^+$ is equivalent to that between $\alpha^+$ and $\beta^-$ as identified by $m_z$ symmetry operation.
If $\beta^-$ is replaced by $\gamma^-$, however, the DWs are reversed as shown in Fig.~\ref{DW_definition}(d).

First principles spin-polarized calculations were done with projector-augmented wave method~\cite{PhysRevB.50.17953} as implemented in {\sc VASP}~\cite{PhysRevB.54.11169}. 
The exchange-correlation interactions among electrons were treated by the local density approximation with Hubbard $U$ correction~\cite{PhysRevB.23.5048,PhysRevB.57.1505}.
The parameters were set to the values of previous reports ($U=8$ and $J=0.88$~eV on the Mn-3$d$ orbitals)~\cite{JPhysCondensMatter.12.4947}, and $A$-type magnetic configurations are adopted.
Lattice constants for the calculation of DWs were fixed to the values of the relaxed unit cell without DWs, and internal positions were optimized in each case 
until the forces acting on all atoms converged to less than 0.005~eV/\AA\ respectively. 
We carefully tested the convergence of the plane wave cutoff energy and $k$-point sampling. 
The cell size dependence was checked by adopting several supercells containing up to 300 atoms.

\begin{figure*}
  \includegraphics[width=1\linewidth]{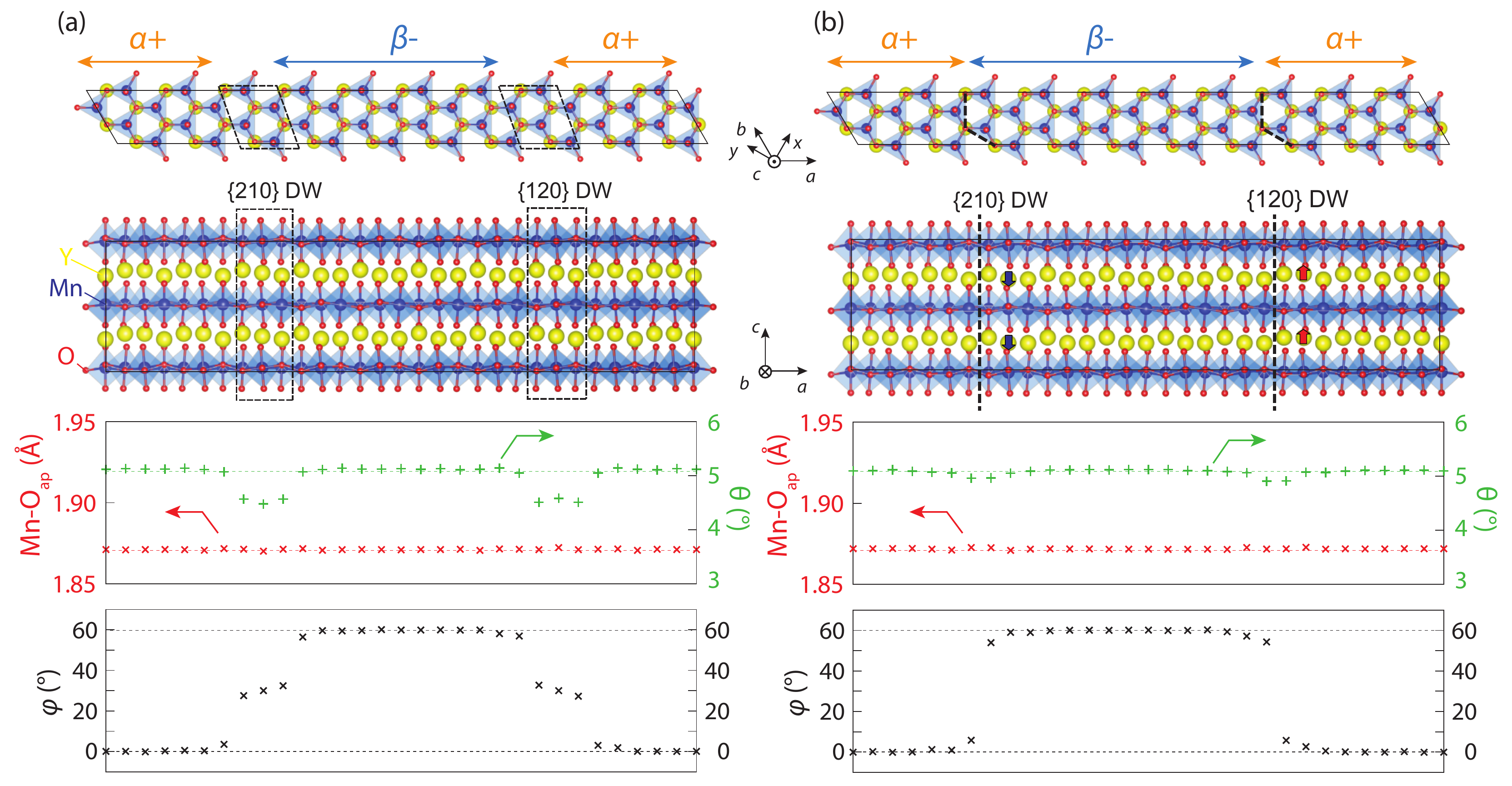}
  \caption{Calculated $\{120\}$ and $\{210\}$ DW structures for (a) the $2/c$ DW models and (b) the sharp models (see text).
           The average Mn-apical O distance, and tilting magnitude ($\theta$) and azimuthal angle ($\varphi$) of MnO$_5$ bipyramids for each layer are also shown. 
           Dashed lines indicate the bulk quantities, which are recovered at the middle of each domain.
           Two-fold rotation and $c$-glide reflection symmetries exist in the $2/c$ DW models whereas two-fold screw rotation and $c$-glide reflection symmetries exist in the sharp models.
           The uniform Mn-O$_{\rm ap}$ distances even through the DWs indicates MnO$_5$ bipyramids rigidly tilt even through the DWs.
}
  \label{DW_models}
\end{figure*}

We begin by calculating the energies of the 
two DW models originally proposed by Choi $et~al$. in Ref.~\cite{NatMat.9.253}. These have the orientation of 
the $\{120\}$ and $\{210\}$ DWs in our nomenclature, and are obtained by averaging the structures of 
two neighboring domains so that the Y ions along the wall retain their high symmetry positions, and the DWs retain
two-fold rotation and $c$-glide reflection symmetries. Therefore we refer to them as the $2/c$ models.
These $2/c$ models are physically reasonable since walls that are the structural average of two different
domains, and which therefore retain aspects of the paraelectric structure, 
are known to be stable in perovskite-structure ferroelectrics~\cite{PhysRevB.65.104111,PhysRevB.80.104110,NatMat.7.57}.
Figure~\ref{DW_models}(a) shows our calculated lowest energy walls within the constraints of the $2/c$ symmetry for $\{120\}$ 
and $\{210\}$ DWs in a 300-atom supercell. First, we note that, after structural relaxation, only the constrained Y atom 
directly at the boundary adopts its paraelectric position; on either side the Y ions immediately adopt their bulk positions,
indicating that the DW is only one unit cell (u.c.) wide. 
Second, we note that the intermediate structure at the DW has $\varphi\approx30^{\circ}$. This is reminiscent of the
centrosymmetric $P\overline{3}c$ InMnO$_3$ structure in which $\frac{1}{3}$ of the In ions retain six-coordinated 
high-symmetry positions and the MnO$_5$ bipyramids trimerize at angles intermediate to those of the YMnO$_3$
structure ($\varphi=30^{\circ}+n\cdot60^{\circ}$)~\cite{PhysRevB.85.174422}.
We find also that the tilting magnitudes of the MnO$_5$ bipyramids $\theta$ are lower at the DWs 
since the tilting correlate with the displacement of Y ions~\cite{PhysRevB.72.100103} and $\frac{1}{3}$ of Y ions keep staying at paraelectric positions at the DWs.
Indeed, the tilting magnitudes of YMnO$_3$ in $P6_3cm$ and $P\overline{3}c$ structures are respectively 5.1 and 4.5 degrees.
The $2/c$ DW energy, calculated with four supercells then linearly extrapolated to infinite spacing 
yields 390 meV/supercell = 195 meV/DW = 195 meV/($a \cdot c$) $\approx$ 44.8 mJ/m$^2$. 
This is close to the energy difference per unit cell between the $P6_3cm$ and $P\overline{3}c$ structures (155 meV/u.c.)~\cite{PhysRevB.85.174422},
consistent with the fact that the width of each DW is approximately 1 u.c. of $P\overline{3}c$ structure.
The slightly higher energy would come from the walls between the $P6_3cm$ and $P\overline{3}c$ structures.

\begin{figure}[b]
  \includegraphics[width=1\linewidth]{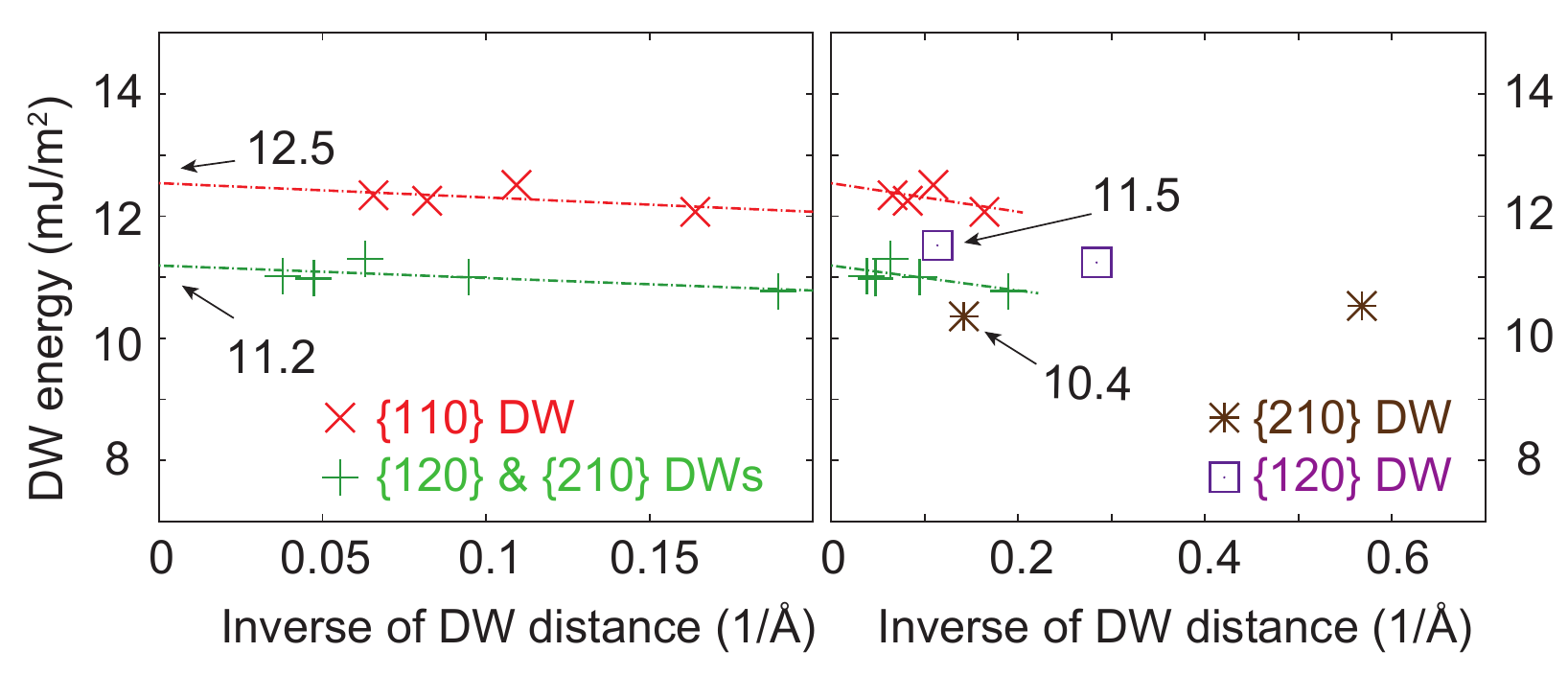}
  \caption{Energies for $\{110\}$, $\{120\}$ and $\{210\}$ DWs calculated with different supercell sizes. 
           The dash-dotted lines are linear extrapolations to infinite spacing (zero inverse distance). 
           Plus signs ($+$) indicate the averages of $\{120\}$ and $\{210\}$ DW energies calculated with the supercells containing both DWs. 
           As expected, these values correspond approximately to the average of the individual $\{120\}$ and $\{210\}$ DW energies calculated with the large supercells containing one type of DW each.}
  \label{DW_energy}
\end{figure}

\begin{figure}
  \includegraphics[width=1\linewidth]{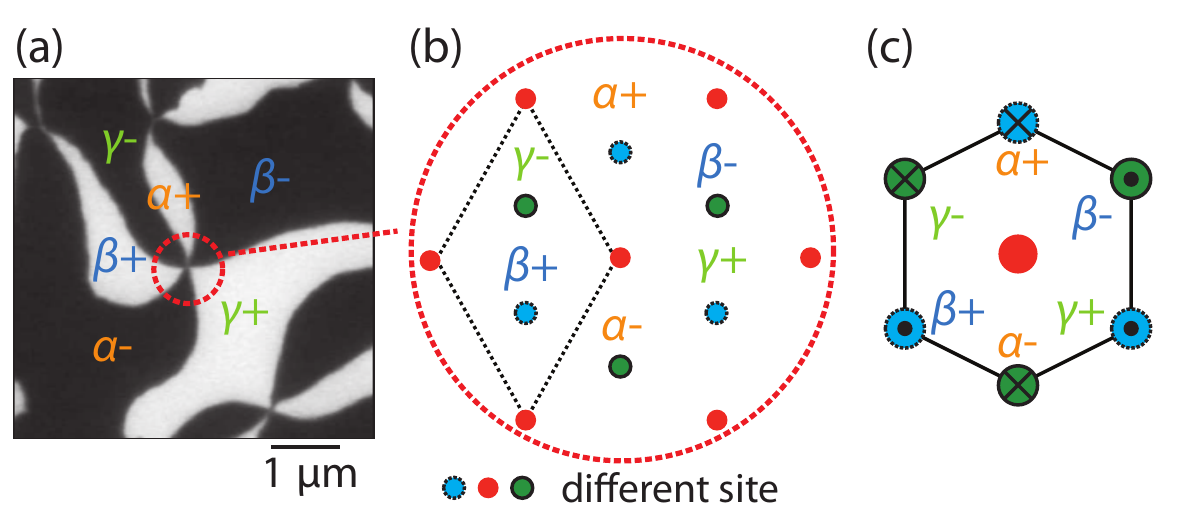}
  \caption{(a) Experimentally observed domain pattern with PFM reproduced from Ref.~\onlinecite{PhysRevB.85.174422} and 
           (b--c) pictorial view of three Y sites and their possible displacements around the topological defect.
           Note that around the center, $\alpha^{+ (-)}$, $\beta^{+ (-)}$, and $\gamma^{+ (-)}$ share the same site, 
           and thus two of them displace up (down) and one of them down (up).}
  \label{topological_defect}
\end{figure}

Next, we displace the centrosymmetric Y ions at the DWs along the $z$ axis as shown in Fig.~\ref{DW_models}(b) and fully relax the internal 
positions, without the constraint of the symmetry.
Our subsequent structural optimization does not recover the $2/c$ DW models and instead yields lower energy DWs in which two-fold rotation symmetry breaks and 
two-fold screw rotation symmetry appears.
In these low energy walls, $\varphi$ and $\theta$ change abruptly across the DW from one bulk value to the next, and 
we can consider the width of $\{120\}$ and $\{210\}$ DWs in YMnO$_3$ to be effectively zero; we refer to them subsequently
as sharp DWs. We note that our calculated $\{210\}$ DW is consistent with recent STEM measurement of the Tm positions in hexagonal
TmMnO$_3$ for walls within the bulk of the samples~\cite{PhysRevB.85.020102}.
It is worth pointing out that $\{210\}$ DW for the $2/c$ model is observed at the edge of the samples in TmMnO$_3$.
This may be due to the truncation of the electrostatic potential or defect accumulation at the edge of the sample.
The wall width is much thinner than expected for a ferroelectric DW, which is typically a few u.c. wide, but rather characteristic for an antiphase boundary. 
The DW energy does not strongly depend on the supercell size (Fig.~\ref{DW_energy}), indicating that the adjacent walls are neither
strongly repulsive nor attractive. Our calculated DW energy extrapolated to infinite spacing is 11.2 mJ/m$^2$, which is a quarter of the 
value calculated for the $2/c$ DWs.
Microscopically, the DWs intersect the shared edges of the MnO$_5$ bipyramids and are staggered when viewed from $c$-axis (Fig.~\ref{DW_models}(b)).
The length is $2/\sqrt{3}=1.15$ times longer than that of the straight line observed in mesoscopic experiments such as piezoresponce force microscopy (PFM) (see also Fig.~\ref{topological_defect}(a)).
In $R$MnO$_3$ polar $\Gamma_2$ mode, i.e. ferroelectric polarization couples to the $K_3$ mode 
and the polarization is proportional to the amplitude of $K_3$ mode $Q_{K_3}$ at $Q_{K_3} > 0.5 $~\cite{PhysRevB.72.100103}.
Since $\theta$ is almost constant even close to the DWs and thus $Q_{K_3} \approx 1$, the polarization direction should abruptly reverse across the DWs with similar bulk polarization.

\begin{figure*}
  \includegraphics[width=1\linewidth]{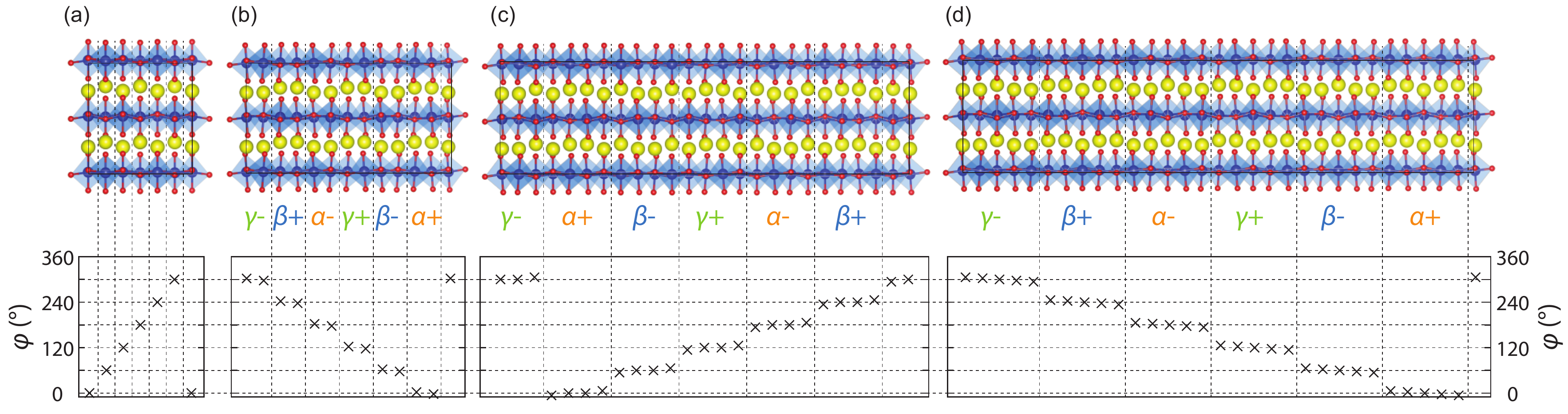}
  \caption{Calculated supercells for evaludating the $\{210\}$ and $\{120\}$ AP+FE DW energies as well as structural phases $\varphi$ for each layer. 
           (a) 60- and (c) 240-atom supercells contain six $\{210\}$ DWs and (b) 120- and (d) 300-atom supercells contain six $\{210\}$ DWs so as to retain the periodic boundary condition.
           In all cases the DWs are evenly spaced.
}
  \label{6DWs}
\end{figure*}

The supercells considered above each contain both $\{120\}$ and $\{210\}$ DWs, and the calculated DW energy is the average value. 
To extract individual DW energies, we next construct supercells containing only one type of DW.
Since the structural phase $\varphi$ shifts $+60^{\circ}$ and $-60^{\circ}$ in the [210] direction through $\{210\}$ and $\{120\}$ DWs, 
respectively, a supercell containing six identical DWs is required to retain the periodic boundary condition.
We use 60- and 240-atom supercells for $\{210\}$ DW, and 120- and 300-atom supercells for $\{120\}$ DW; in all cases the DWs are evenly spaced (Fig.~\ref{6DWs}).
Figure~\ref{DW_energy}(b) shows our calculated $\{210\}$ and $\{120\}$ DW energies from the two different supercells.
Surprisingly, although the distances between DWs in 60- and 120-atom supercells are only $\frac{\sqrt{3}}{2}\cdot\frac{1}{3}a$ and $\frac{\sqrt{3}}{2}\cdot\frac{2}{3}a$, respectively,
the estimated DW energies are very close to those with 240- and 300-atom supercells.
This indicates that interactions between the DWs are very weak.
We find that the energy of the $\{120\}$ DW is slightly higher than that of $\{210\}$ DW.
The Y ions displace up-down-up-down across $\{210\}$ DWs and up-up-down-down across $\{120\}$ DWs (Fig.~\ref{DW_models}), 
which likely contributes to differences in the electrostatic energies. 

\begin{figure}
  \includegraphics[width=1\linewidth]{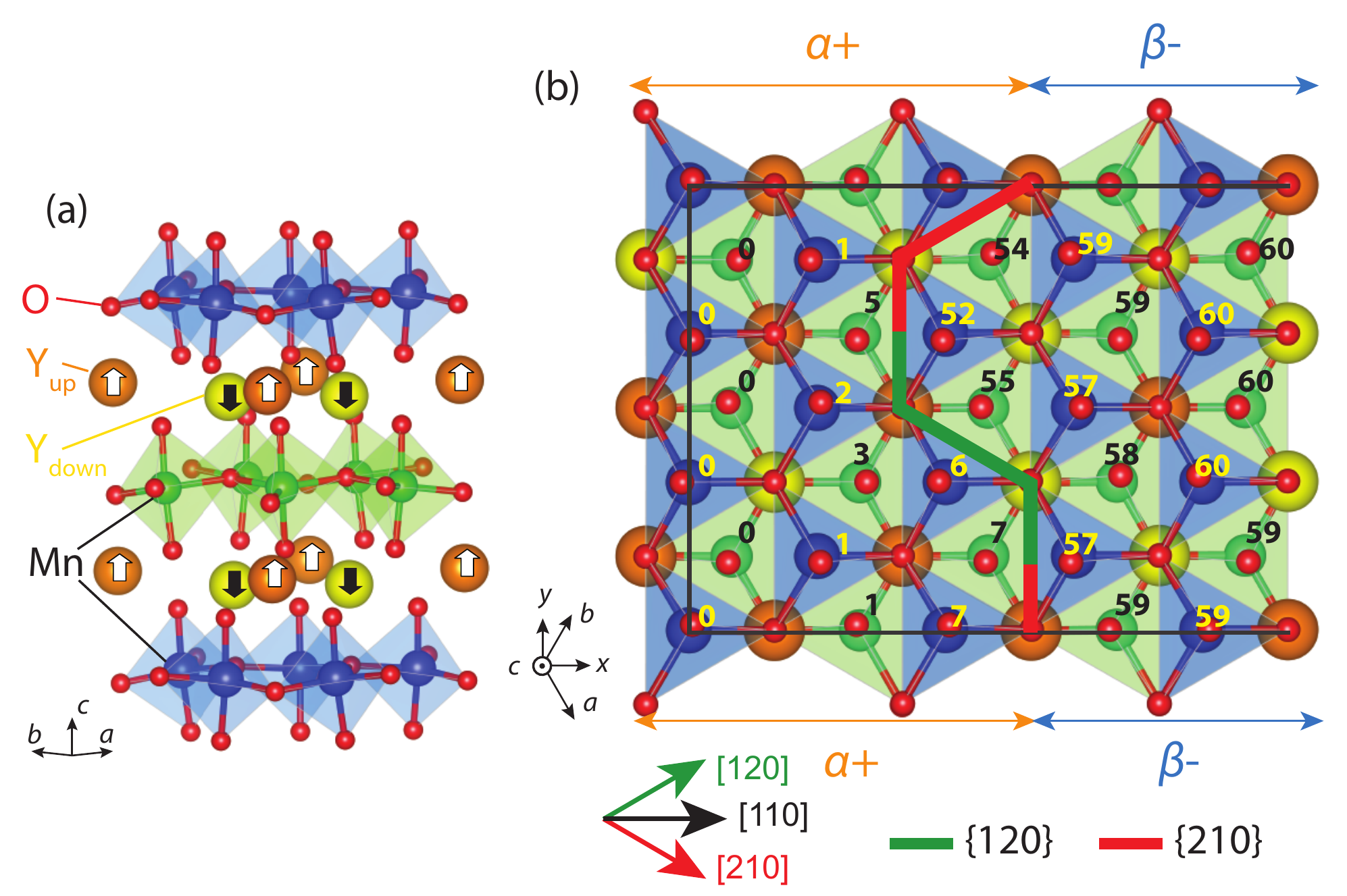}
  \caption{(a) Unit cell of YMnO$_3$. Mn ions staying at different layer and Y ions displacing in the opposite direction are distinguished by different colors.
           (b) Calculated $\{110\}$ DW structure for the sharp model with a 300-atom supercell. 
           Only half of the supercell is shown with consideration for the supercell symmetries.
           Numbers indicate the average structural phases $\varphi$ at each column of MnO$_5$ bipyramids.
           The thick line indicates the $\{110\}$ DW derived from the phase shift.
           Note that the $\{110\}$ DW is regarded as a combination of $\{120\}$ and $\{210\}$ DWs.
}
  \label{110_DW_model}
\end{figure}

Next we calculate the structure and energetics of $\{110\}$ DWs.
Figure~\ref{110_DW_model} shows our relaxed $\{110\}$ DW structure calculated within a 300-atom supercell.
Based on our results for $\{120\}$ and $\{210\}$ DWs, we constructed the initial structure by displacing the Y ions along the $z$ axis so that domains suddenly change through the DWs.
This sharp configuration is retained on fully relaxing the supercell, and our calculated DW width is effectively zero. 
We note that two-fold screw rotation and $c$-glide reflection symmetries exist in the supercells for this wall orientation, and so the $\{110\}$ and $\{\bar{1}\bar{1}0\}$ DWs 
are symmetry-equivalent and we can extract their energies directly from the supercells containing both walls.
Our calculated infinite-separation DW energy extrapolated from calculations for four supercells is 12.5 mJ/m$^2$ (Fig.~\ref{DW_energy}), which
is slightly higher than those of $\{120\}$ and $\{210\}$ DWs
In fact the $\{110\}$ DW can be regarded as a combination of $\{120\}$ and $\{210\}$ DWs (Fig.~\ref{110_DW_model}),
and indeed the $\{110\}$ DW energy estimated from the $\{120\}$ and $\{210\}$ DW energies is $2/\sqrt3\cdot11.2=12.9$ mJ/m$^2$ 
which is close to our explicitly calculated value.
We note also that the length of the $\{110\}$ DW viewed from $c$ axis is $4/3=1.33$ times longer than that of the straight line.

\begin{figure}
  \includegraphics[width=1\linewidth]{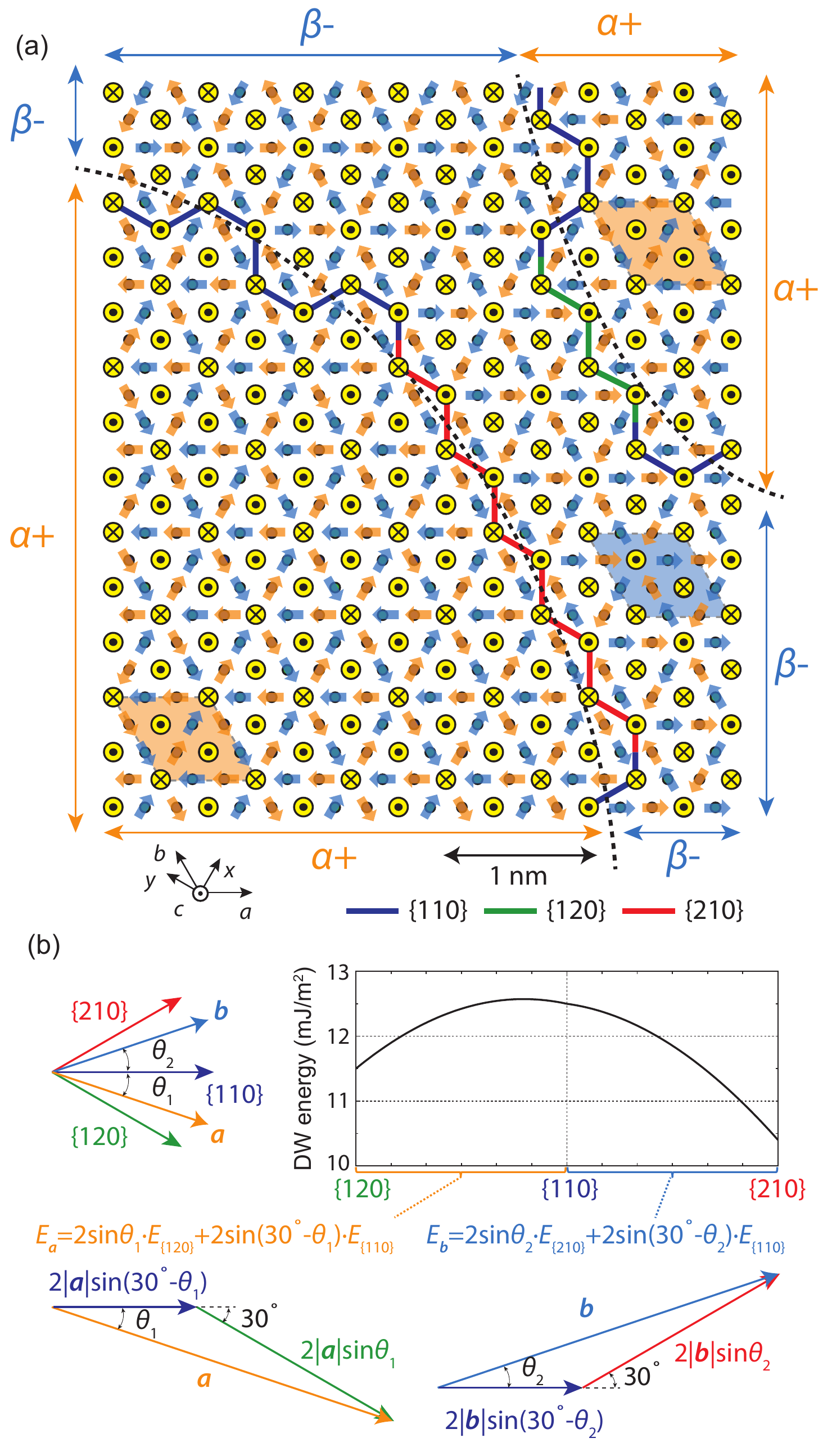}
  \caption{(a) Schematics of winding AP+FE DWs, which are combinations of $\{120\}$, $\{210\}$, and $\{110\}$ DWs.
           The dashed lines indicate the mesoscopically observed DWs and the directions of Y ion displacements are changed through the DWs. 
           The thick lines indicate the  DWs derived from the phase shift.
           (b) DW energy as a function of the DW direction, which is geometrically interpolated with $\{120\}$, $\{210\}$, and $\{110\}$ DW energies represented as $E_{\{120\}}$, $E_{\{210\}}$, and $E_{\{110\}}$, respectively.
             }
  \label{general_DW}
\end{figure}

Our finding that the DW width is almost zero is in contrast to the behavior in conventional perovskite counterparts 
such as PbTiO$_3$, where the widths of 180$^\circ$ DWs are narrow but finite ($\geq1$ u.c. width)~\cite{PhysRevB.65.104111,PhysRevB.80.104110,NatMat.7.57},
and the ions in the wall region transition from one orientation to the other through their high-symmetry paraelectric positions. The adoption
of the high-symmetry position can be regarded as a geometric frustration of the positions in the two opposite domain structures. 
This frustration does not exist in h-$R$MnO$_3$, because the up-up-down or down-down-up displacements of the $R$ ions, determined by the tiltings of the MnO$_5$ trigonal bipyramids,
can be locally preserved through the DWs, as seen in Figs~\ref{DW_models} and \ref{110_DW_model}.
Instead, the translational periodicity that is broken at the DWs is reminscent of
${\it stacking}$ ${\it faults}$ commonly observed in close-packed structures.
It is also interesting to note that the 180$^\circ$ DW energy in YMnO$_3$ is considerably lower than those of perovskites with similar 
Curie temperatures, for example the corresponding 180$^\circ$ DW energy in PbTiO$_3$ ($T_{\rm C}$=765 K) is
132 mJ/m$^2$~\cite{PhysRevB.65.104111} and in BiFeO$_3$ ($T_{\rm C}$=1100 K)
it is 829 mJ/m$^2$~\cite{PhysRevB.80.104110}. While in conventional ferroelectrics, Curie temperatures often correlate with the
magnitude of the ferroelectic polarization, this is not the case in 
YMnO$_3$ where the Curie temperature is high (1258$\pm$14 K)~\cite{PhysRevB.83.094111} and corresponds to the tilting transition~\cite{PhysRevB.72.100103},
while the improper ferroelectric polarization is not large (5.6 $\mu$C/cm$^2$). 

Experimentally, most DWs do not form straight lines corresponding to a single orientation, but wind, as illustrated in Fig.~\ref{general_DW}
for the case of our calculated lowest energy sharp walls.
It is clear from Fig.~\ref{general_DW} (a) that the winding DWs consist of combinations of $\{120\}$ (green), $\{210\}$ (red), and $\{110\}$ (blue) DWs, 
which in turn are combinations of $\{120\}$ and $\{210\}$ DWs.
We can then estimate the energy of a DW pointing in arbitrary direction from geometrical relations. 
Our results are shown in Fig.~\ref{general_DW}(b), where we see that the $\{210\}$ DW has the lowest energy.
This is consistent with the recent observation of stripe domain patterns in h-$R$MnO$_3$ grown below the trimerization temperatures~\cite{PhysRevLett.108.167603}.
The stripes develop along the $[110]$ direction, which could correspond to either $\{120\}$ and/or $\{210\}$ DWs.
From our results, we suggest that the DWs are likely to be of $\{210\}$ type.
Since $\{210\}$ DWs do not vanish when they meet together for the same direction, 
they should lead to a topological protection of the stripe domains under electric-field application even without the topological defects.

\begin{figure}
  \includegraphics[width=1\linewidth]{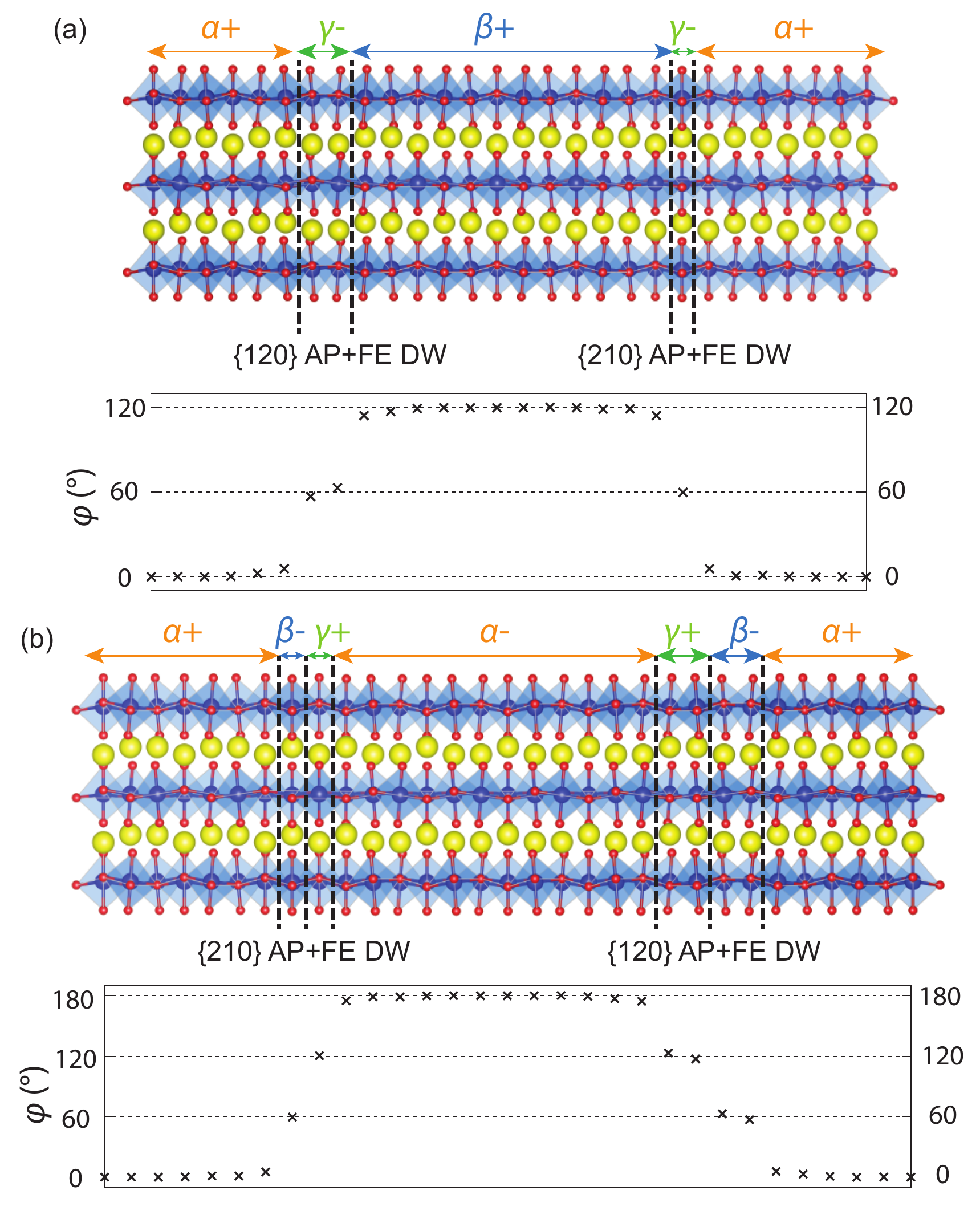}
  \caption{Calculated (a) AP and (b) FE DWs perpendicular to the [210] axis with 270- and 300-atom supercells, respectively. 
           The models locally preserve the up-up-down or down-down-up displacements of the $R$ ions across the DWs, resulting in the interstitial domains.}
  \label{FE_AP_DWs}
\end{figure}

We now perform structural relaxations for FE-only and AP-only DWs, to address the question of why such walls do not occur in h-$R$MnO$_3$.
Based on the results of the AP+FE DWs, we constructed initial models to preserve the up-up-down or down-down-up displacements of the $R$ ions through the DWs.
As shown in Fig.~\ref{FE_AP_DWs}, the relaxed structures retain the off-centering of the Y ions. By analyzing
the tilt patterns of the MnO$_5$ polyhedra, and the corresponding Y off-centering patterns, we see that an AP DW 
in fact is formed from two AP+FE DWs, and a FE DW is composed of three AP+FE DWs. 
Indeed, the average DW energies that we obtain for AP and FE DWs with 270- and 300-atom supercells are 23.0 and 34.4 mJ/m$^2$ 
respectively; almost exactly two and three times the average value for the AP+FE DWs (11.2 mJ/m$^2$).
This is because the phase difference between two domains with different origin (polarization) is $120^{\circ}$ ($180^{\circ}$), whereas 
AP+FE DWs can change the phase only $\pm60^{\circ}$.
The obtained DW energies are not consistent with the six-state clock model used in Ref.~\cite{PhysRevLett.108.167603} for Monte Carlo simulation,
in which the ratio of AP+FE, AP, and FE DW energies is assumed to be 1:3:4.
We also see that the minimum possible width of the intermediate phase is $\frac{\sqrt{3}}{2}\cdot\frac{1}{3}a$ or $\frac{\sqrt{3}}{2}\cdot\frac{2}{3}a$, depending on the type of the 
AP+FE DW. 
Jungk $et~al$. have found a width of 60$\pm$10 nm for the interstitial domain with PFM after applying electric field~\cite{ApplPhysLett.97.012904}.
This would be due to the depolarization field and/or the resolution of PFM.
Finally we note that, in our energy minimizations, the AP and FE DWs do not tend to yield larger
regions of the intermediate phase calculations; this is because, as we discussed earlier, the interaction between
DWs is very weak and so there is not a strong driving force for them to separate. At the high temperatures close
to the phase transition temperature, however, we expect AP+FE DWs to randomly distribute for entropic reasons, and 
the artificially narrow intermediate regions shown in Fig.~\ref{FE_AP_DWs} would naturally broaden so that 
AP-only and FE-only DWs would not survive.  

\begin{figure}
  \includegraphics[width=1\linewidth]{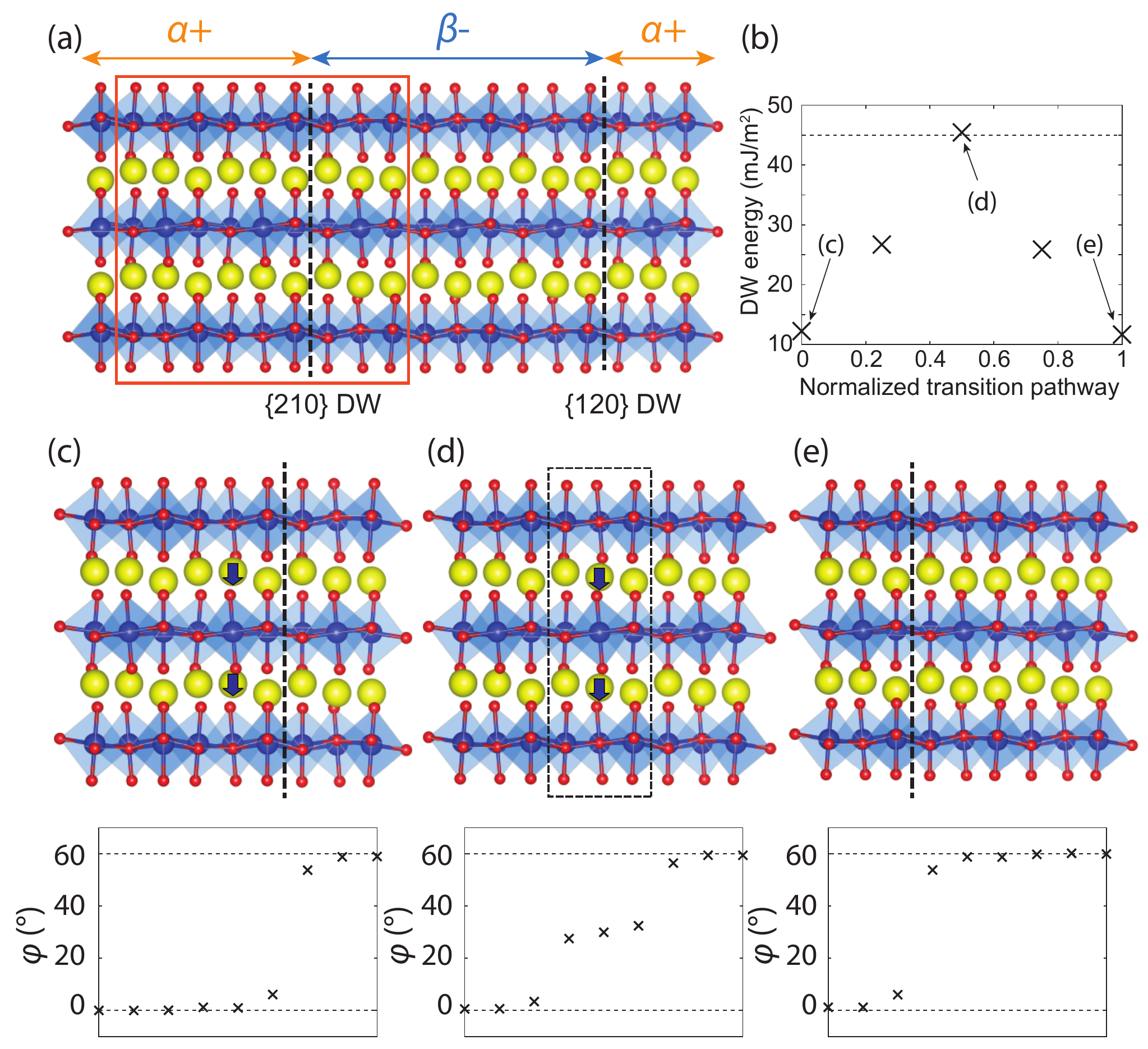}
  \caption{(a) Initial structure for calculating the migration of $\{210\}$ AP+FE DW with NEB method.
           (b) The calculated energy landscape through the DW migration.
           The dashed line indcates the average DW energy of $\{210\}$ and $\{120\}$ $2/c$ DWs.
           (c--e) The initial, intermediate, and final structures in the frame in (a).
           }
  \label{migration}
\end{figure}

In ferroelectrics an electric polarization is reversed by the application of an electric field through the migration of the DW,
and thus it is valuable to understand the microscopic migration mechanism.
In Fig.~\ref{migration} we present the energy landscape through the DW migration and the transition structure calculated with the nudged elastic band (NEB) method~\cite{SurfSci.324.305}.
It is interesting that the transition structure and its energy at the middle point are almost same with those of the $2/c$ DW (Fig.~\ref{DW_models}).
In other word, the $2/c$ DW is the local maximum of the migration energy curve.
This is because when the DW moves, Y ions specified in Fig.~\ref{migration}(c) go through the paraelectric sites, and simultaneously the MnO$_5$ trigonal bipyramids rotate.
We also see that the DW width increases from approximately zero to 1 u.c. in the migration,
and finally moves from one edge of the intermediate $2/c$ DW to another, meaning the minimum migration step is 1 u.c. 
The energy barrier height is 33.2 mJ/m$^2$, and is almost three times of the DW energy.
Although the barrier height is comparable to PbTiO$_3$ (37 mJ/m$^2$), 
the ratio to the DW energy is much higher than that of PbTiO$_3$, of which the energy barrier is only 36\% of the DW energy~\cite{PhysRevB.65.104111}.
However, the energy barrier height would significantly reduce at the high temperatures close to the phase transition temperature, and AP+FE DWs would result in random distribution.

Based on our finding that the lowest energy AP+FE DWs have the sharp configuration, in this last section we discuss the likely structure 
of the topological defects that form at the meeting points between six AP+FE DWs.
Figure~\ref{topological_defect} shows the displacements of Y ions in the AP+FE DWs near the topological defect.
Of the six Y ions around the central point, half of them displace up and half down along the $z$ axis.
This causes a frustration of the Y ion at the center of the topological defect which can not displace in a direction
that allows all DWs to maintain their sharp configuration. 
It is possible, therefore, that the Y ion at the center might remain at its paraelectric position; otherwise a modification
of the wall structure would be required in the vicinity of the topological defect.
We note that this picture does not correspond to that of Ref.~\cite{NatMat.9.253}, as the latter paper used the higher energy
$P2/c$ models for the DWs meeting at the defect. 
It is clear that further theoretical and microscopy studies would be desirable for a full understanding of the structure of 
the topological defects.

We thank Manfred Fiebig and Dennis Meier for fruitful discussions. 
This work was supported by ETH Z\"{u}rich, the European Research Council FP7 Advanced Grants program, grant number 291151 and 
the JSPS Postdoctoral Fellowships for Research Abroad (YK).
The visualization of crystal structures was performed with {\sc vesta}~\cite{JApplCryst.41.653}.

\end{document}